\documentclass[twocolumn, 10pt,journal]{IEEEtran}
\ifCLASSINFOpdf
\else
\fi
\usepackage{mathrsfs}
\usepackage{graphicx}
\usepackage{epsfig}
\usepackage{amsmath}
\usepackage{bm}
\usepackage{float}
\usepackage{multirow}
\usepackage{color}
\usepackage{subfig}
\usepackage{amsmath}
\usepackage{algorithm}
\usepackage{algorithmic}
\usepackage{cite}
\usepackage{amsthm}
\usepackage{amssymb}
\usepackage{tabularx}
\usepackage{booktabs}
\usepackage{color}

\usepackage{CJK}
\usepackage{amsmath}

\newenvironment{sequation*}{\begin{equation*}\small}{\end{equation*}}

\theoremstyle{definition}

\theoremstyle{remark}

\theoremstyle{proposition}

\long\def\symbolfootnote[#1]#2{\begingroup%
\def\thefootnote{\fnsymbol{footnote}}\footnote[#1]{#2}\endgroup}

\ifCLASSOPTIONcompsoc
\usepackage[nocompress]{cite}
\else
\usepackage{cite}
\fi

\begin{document}
\pagestyle{empty}   

\bibliographystyle{IEEEtran}

\title{Efficient Radio Resource Management for Wireless Cellular Networks with Mobile Edge Computing}

\author{Chenmeng Wang\IEEEauthorrefmark{1} and S. Hu\IEEEauthorrefmark{2}\\

\IEEEauthorblockA{\IEEEauthorrefmark{1}Chongqing Key Lab of Mobile Communications Technology,\\
Chongqing University of Posts and Telecomm. (CQUPT), Chongqing, 400065, P.~R.~China}\\
\IEEEauthorblockA{\IEEEauthorrefmark{2}Dept. of Systems and Computer Eng., Carleton University, Ottawa, ON, Canada}\\
}
\maketitle
\thispagestyle{empty}  

\begin{abstract}
Mobile edge computing (MEC) has attracted great interests as a promising approach to augment computational capabilities of mobile devices. An important issue in the MEC paradigm is computation offloading. In this paper, we propose an integrated framework for computation offloading and interference management in wireless cellular networks with mobile edge computing. In this integrated framework,
the MEC server makes the offloading decision according to the local computation overhead estimated by all user equipments (UEs) and the offloading overhead estimated by the MEC server itself. Then, the MEC server performs the PRB allocation using graph coloring. The outcomes of the offloading decision and PRB allocation are then used to allocate the computation resource of the MEC server to the UEs. Simulation results are presented to show the effectiveness of the proposed scheme with different system parameters.

\end{abstract}


\section{Introduction}

As the popularity of smart phones is increasing dramatically, more and more new mobile applications are emerging, such as face recognition, natural language processing, and augmented reality \cite{WD13}. This has led to an exponential growth of demand in not only \emph{high data rate} but also \emph{high computational capability} in wireless cellular networks\cite{ASA14}.

One recently proposed solution for addressing the  data rate issue is the use of small cells \cite{HM12, XYJL12,BY14}.
Nevertheless, there exists
inter-cell interference, which  can significantly degrade network performance. Without proper \emph{interference management},
 the overall spectrum efficiency and energy efficiency of the network might become even worse than that of a network without small cells \cite{Interference2015Huang,BYY15}.

On the other hand, to address the  computational capability issue, \emph{mobile edge computing} (MEC), is being standardized to allocate computing resources  in wireless cellular networks \cite{MEC}. MEC allows mobile user equipments (UEs) to perform \emph{computation offloading} to offload their computational tasks to the MEC server via wireless cellular networks. Then each UE is associated with a clone in MEC server, which executes the computational tasks on behalf of that UE.

Although some outstanding works have been dedicated in studying computation offloading and interference management, these two important aspects were generally considered separately in the existing works.
Therefore, in this paper, we propose to jointly consider computation offloading and interference management in order to improve the performance of wireless cellular networks with mobile edge computing. The motivations behind our work are based on the following observations.

\begin{itemize}
\item 
The experience of end-to-end application (e.g., video) indicates that the optimized performance in one segment of the whole system does not guarantee the end-to-end user experience \cite{CW12}.

\item If multiple UEs choose to offload their computational tasks to the MEC server via small cell networks simultaneously, severe interference can be generated. Moreover, the MEC server could be overloaded. So, some UEs should be selected to offload their computations, while others should execute their computations locally.
\end{itemize}

The distinct features of this paper are as follows.

\begin{itemize}
\item We propose an integrated framework for computation offloading and interference management in wireless cellular networks with mobile edge computing. 

\item In the framework, the MEC server makes the offloading decision according to the estimated system overall overhead. Based on the offloading decision, the MEC server then performs the PRB allocation using graph coloring. The outcomes of the offloading decision and PRB allocation are then used to allocate the computation resource of the MEC server to the UEs.

\item Simulation results are presented to show the effectiveness of the proposed scheme with different system parameters.

\end{itemize}


The rest of this paper is organized as follows. The system model under consideration is described in Section \ref{Sec 2: system model}. The proposed integrated framework is presented in Section \ref{Sec 3: building blocks}. Simulation results are discussed in Section \ref{numerical results}. Finally, we conclude this study in Section \ref{conclusion}.


\section{System Model}
\label{Sec 2: system model}

\subsection{Network Model}
An MEC server is placed in the macro eNodeB (MeNB), and all the $N$ small cell eNodeBs (SeNBs) are connected to the MeNB as well as the MEC server. The set of small cells is denoted by $\mathscr{N}=\{1,2,...,N\}$. For the sake of simplicity, it is assumed that each SeNB $n$ is associated with only one mobile UE. We assume that UE $n$ is associated with SeNB $n$.
 We assume that each UE has a
task to be completed. Each UE could offload the computation to the MEC server through the SeNB with which it is associated, or execute the computation task locally. For simplicity, user mobility and handover \cite{MYL04,YL01,YK07} are not considered. 

\subsection{Communication Model}
\label{communication model}
We denote $a_n\in\{0,1\}$ as the computation offloading decision of UE $n$. Specifically,
We have $a_n=1$ if UE $n$ chooses to offload the computation to the MEC server via wireless access, and $a_n=0$ otherwise. So we have $\boldsymbol{A}=\{a_1, a_2,... a_N\}$ as the offloading decision profile.

In this paper, we consider uplink direction where transmission is from a UE to the associated SeNB, and interference is from a UE to a neighboring SeNB.
We denote the total number of physical resource blocks (PRBs) as $K$. Here we introduce a PRB association table $\boldsymbol{C}$, which is an $N \times K$ table with binary entries $c_{n,k}$, where $N$ is the total number of SeNBs and $K$ is the total number of PRBs. The entry $c_{n,k}$ in the association table is set to 1 if SeNB $n$ is assigned with PRB $k$ and 0 otherwise. Given the decision profile $\boldsymbol{A}=(a_1,a_2,...,a_N)$ and the PRB association table $\boldsymbol{C}$, the uploading rate achieved by UE $n$ connected to SeNB $n$ is given by
\begin{equation}
\setlength{\abovedisplayskip}{2pt}
\setlength{\belowdisplayskip}{2pt}
\begin{aligned}
\label{rate achieved by n}
R_n(\boldsymbol{A},\boldsymbol{C})&= a_n \cdot \sum\limits_{k=1}^{K} c_{n,k} \cdot \frac{B}{K} \cdot\\
& \log_{2}\left(1+\frac{\frac{P_n}{M_n}H_{n,n}}{\sigma^2+\sum\limits_{m=1,m \neq n}^{N} a_m \cdot c_{m,k} \cdot \frac{P_m}{M_m} H_{m,n}}\right),
\end{aligned}
\end{equation}
where $P_n$ denotes the transmission power of UE $n$, $\sigma^2$ denotes noise variance per PRB, $M_n$ stands for the number of PRBs assigned to small cell $n$, and $H_{n,n}$, $H_{m,n}$ stand for the channel gain between UE $n$ and SeNB $n$, the channel gain between UE $m$ and SeNB $n$, respectively.


\subsection{Computation Model}
For the computation model, we consider that each UE $n$ has a computation task $I_n \triangleq (B_n, D_n)$. Here $B_n$ stands for the size of input data, while $D_n$ denotes the total number of CPU cycles required to accomplish the task.

\subsubsection{Local Computing}
\label{local computing}
For the local computing approach, the computation task $I_n$ is executed locally on each mobile device. We denote $F_n^{(l)}$ as the computational capability (i.e., CPU cycles per second) of UE $n$.
The computation execution time $T_n^{(l)}$ of task $I_n$ executed locally by UE $n$ is expressed as
\begin{equation}
\label{local computation time}
T_n^{(l)}=\frac{D_n}{F_n^{(l)}},
\end{equation}
and the computational energy consumption $E_n^{(l)}$ is given by
\begin{equation}
\label{local computation energy}
E_n^{(l)}=v_nD_n,
\end{equation}
where $v_n$ is the coefficient representing the energy consumed by each CPU cycle. According to the realistic measurements in \cite{wen2012energy}, we set $v_n=10^{-11}(F_n^{(l)})^2$.

According to (\ref{local computation time}) and (\ref{local computation energy}), the total overhead of the local computing approach on UE $n$, in terms of computational time and energy, $Z_n^{(l)}$ can be calculated as
\begin{equation}
\label{local total overhead}
Z_n^{(l)}=\gamma_n^{(T)}T_n^{(l)}+\gamma_n^{(E)}E_n^{(l)},
\end{equation}
where $0\leq \gamma_n^{(T)}, \gamma_n^{(E)} \leq 1$ represent the weights of computational time and energy of UE $n$, respectively.

\subsubsection{MEC Server Computing}
According to the communication model presented in Subsection \ref{communication model}, the time and energy costs for transmitting the computation input data of size $B_n$ are calculated as, respectively,
\begin{equation}
\label{MEC offloading time}
T_{n,off}^{(e)}(\boldsymbol{A},\boldsymbol{C})=\frac{B_n}{R_n(\boldsymbol{A},\boldsymbol{C})},
\end{equation}
and
\begin{equation}
\label{MEC offloading energe}
E_{n,off}^{(e)}(\boldsymbol{A},\boldsymbol{C})=\frac{P_nB_n}{R_n(\boldsymbol{A},\boldsymbol{C})}.
\end{equation}

The MEC server would execute the computation task after offloading. Let $F_n^{(e)}$ denote the computational capability (i.e., CPU cycles per second) of the MEC server assigned to UE $n$. Then the execution time of the MEC server on task $I_n$ is given as
\begin{equation}
\label{MEC execution time}
T_{n,exe}^{(e)}=\frac{D_n}{F_n^{(e)}}.
\end{equation}

According to \cite{xu2015decentralized} and \cite{offloadornot2013barbera}, the total overhead of MEC computing approach in terms of execution time and energy is computed as
\begin{equation}
\label{MEC computation total overhead}
Z_n^{(e)}(\boldsymbol{A},\boldsymbol{C})=\gamma_n^{（T）}[T_{n,off}^{(e)}(\boldsymbol{A},\boldsymbol{C})+T_{n,exe}^{(e)}]+\gamma_n^{（E）}E_{n,off}^{(e)}(\boldsymbol{A},\boldsymbol{C}).
\end{equation}

Like studies in \cite{xu2015decentralized}, the time consumption of computation outcome transmission from the MEC server to UE $n$ is neglected in this work, due to the fact that the size of computation outcome data in general is much smaller than that of the computation input data including the mobile system settings, program codes and input parameters.

\section{The Integrated Framework for Computation Offloading and Interference Management}
\label{Sec 3: building blocks}
\begin{figure}[!t]
\centering
\includegraphics[width=0.48\textwidth]{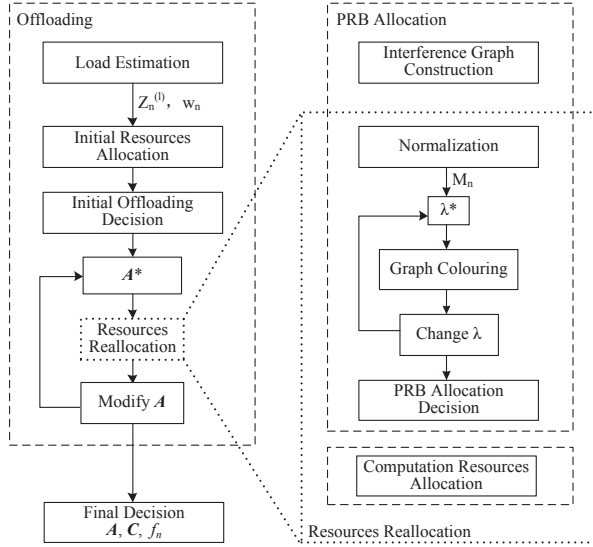}

\caption{Block diagram of the proposed solution.}
\label{fig: block diagram of the proposed solution}
\end{figure}

In this section, we propose an integrated framework for computation offloading and interference management.
We present a suboptimal centralized solution on the MEC server, which
is shown in Fig. \ref{fig: block diagram of the proposed solution}. 

\subsection{Load Estimation}
The overall overhead of local computation can be computed as in (\ref{local total overhead}).
We calculate the minimum number of requested PRBs of UE $n$, $w_n$, as follows,
\begin{equation}
\label{1opt}
\begin{aligned}
\text{minimize} \quad &w_n           \\
\text{subject to} \quad    & C1: w_n \frac{B}{K} \log_{2}\left(1+\frac{P_n H_{n,n}}{w_n \sigma^2}\right)\geq \overline{r_n}    \\
                       & C2: w_n\leq K   \\
                       & C3: w_n \geq 0
\end{aligned}
\end{equation}

The optimization object in (\ref{1opt}) is $w_n$, which represents the minimum number of PRBs required by UE $n$. The first set of constraints $C1$ guarantees that the PRBs assigned to UE $n$ could meet the minimum rate requirement, $\overline{r_n}$. 

The  minimum rate $\overline{r_n}$ is determined by the following steps: 
the time consumption of offloading computation for UE $n$ can be given as,
\begin{equation}
\label{MEC execution time in estimation}
\tilde{T}_{n,exe}^{(e)}=\frac{D_n}{F/N}, \quad  \forall n.
\end{equation}
Then the minimum offloading rate of UE $n$ should be
\begin{equation}
\label{minimum offloading time}
\overline{r_n}=\frac{B_n}{T_n^{(l)}-\tilde{T}_{n,exe}^{(e)}} \quad  \forall n.
\end{equation}

At the end of load estimation, the overall overhead of local computation $Z_n^{(l)}$ and the minimum number of PRBs required for offloading computation $w_n$ is sent to the MEC server.

\subsection{Initial Resource Allocation}
It is obvious that the sum of $w_n$ is not necessarily equal to the total number of PRBs, $K$. So the MEC server then normalizes the loads estimated by UEs as follows:
\begin{equation}
\label{normalization}
\tilde{M}_n=K\frac{w_n}{\sum\limits_{n}w_n},  \quad  \forall n.
\end{equation}
Here we suppose that all UEs will offload computation tasks to the MEC server and the PRBs will be assigned to the UEs in the manner of orthogonal frequency allocation. So UE $n$ will be assigned with $\lfloor \tilde{M}_n\rfloor$ PRBs, and the offloading rate of UE $n$ can be calculated as:
\begin{equation}
\label{initial offloading rate}
\tilde{R}_n=\tilde{M}_n\frac{B}{K}\log_{2}\left(1+\frac{P_nH_{n,n}}{\tilde{M}_n\sigma^2}\right), \quad \forall n.
\end{equation}
Then the time and energy consumption of offloading data for UE $n$ can be given respectively as:
\begin{equation}
\label{initial offloading time}
\tilde{T}_{n,off}^{(e)}=B_n/\tilde{R}_n, \quad \forall n,
\end{equation}
and
\begin{equation}
\label{initial offloading energy}
\tilde{E}_n^{(e)}=P_nB_n/\tilde{R}_n, \quad \forall n.
\end{equation}
The total time consumption of UE $n$ is
\begin{equation}
\label{initial total time}
\tilde{T}_{n}^{(e)}=\tilde{T}_{n,off}^{(e)}+\tilde{T}_{n,exe}^{(e)}, \quad \forall n.
\end{equation}
So the overall overhead of UE $n$ after initial resource allocation is
\begin{equation}
\label{initial total overhead}
\tilde{Z}_{n}^{(e)}=\gamma_n^{(T)}\tilde{T}_{n}^{(e)}+\gamma_n^{(E)}\tilde{E}_{n}^{(e)}, \quad \forall n.
\end{equation}

\subsection{Initial Offloading Decision}

Then the MEC server makes the initial offloading decision for UE $n$ on comparison of its local and offloading computation overhead, i.e., comparison of  $Z_n^{(l)}$ and $\tilde{Z}_{n}^{(e)}$:
\begin{equation}
\label{initial offloading decision}
\left\{
\begin{aligned}
a_n=1, \quad \text{if} \; Z_n^{(l)} > \tilde{Z}_{n}^{(e)}; \\
 a_n=0, \quad \text{if} \;  Z_n^{(l)} \leq \tilde{Z}_{n}^{(e)}. \\
\end{aligned}
\right.
\end{equation}

We assume that the number of non-zero elements in offloading decision vector $\boldsymbol{A}$ is represented by $N_e$, and let $N_l=N-N_e$ represent the number of zero elements in offloading decision vector $\boldsymbol{A}$. Further more, the set of offloading UEs are denoted by $\mathscr{N}_e$, and the set of UEs which will compute locally is denoted by $\mathscr{N}_l$. Then the PRBs are reallocated as:
\begin{equation}
\label{normalization1}
\tilde{M}_n'=K\frac{a_nw_n}{\sum\limits_{n}a_nw_n},  \quad  \forall n,
\end{equation}
where $\tilde{M}_n'$ denotes the number of PRBs reallocated to UE $n$. If $\tilde{M}_n'=0$, $n\in \mathscr{N}_l$; if $\tilde{M}_n'>0$, $n\in \mathscr{N}_e$.

Then, the offloading rate $\tilde{R}_n'$, offloading time $\tilde{T'}_{n,off}^{(e)}$, offloading energy consumption $\tilde{E'}_n^{(e)}$, execution time $\tilde{T'}_{n,exe}^{(e)}$ total time consumption $\tilde{T'}_{n}^{(e)}$ are recalculated using the method in accordance with (\ref{initial offloading rate}) (\ref{initial offloading time}) (\ref{initial offloading energy}) (\ref{MEC execution time in estimation}) and (\ref{initial total time}), respectively.




Then we have the total time consumption and overall overhead of UE $n\in\mathscr{N}_e$ as:
%
\begin{equation}
\label{initial total overhead1}
\tilde{Z'}_{n}^{(e)}=\gamma_n^{(T)}\tilde{T'}_{n}^{(e)}+\gamma_n^{(E)}\tilde{E'}_{n}^{(e)}, \quad n\in \mathscr{N}_e.
\end{equation}
Then we calculate the total overhead of the system as:
\begin{equation}
\label{total system overhead}
Z'=\sum\limits_{i\in\mathscr{N}_l}Z_i^{(l)}+\sum\limits_{j\in\mathscr{N}_e} \tilde{Z'}_{j}^{(e)}.
\end{equation}

\subsection{Resource Reallocation}
\label{Resource Reallocation}
In order to find the optimal offloading decision vector $\boldsymbol{A}^{\ast}$, the initial offloading decision vector $\boldsymbol{A}$ drawn from last subsection is modified and the PRBs and computational resource are reallocated
following the method below:

\begin{itemize}
\item Check the offloading vector $\boldsymbol{A}$. If every element of $\boldsymbol{A}$ equals to 1, then $\boldsymbol{A}$ will be the final offloading decision; if not, follow the steps below:
\item Check the zero elements of offloading profile $\boldsymbol{A}$, and search for the UE $\hat{n}$ with the lowest offloading computation overhead $\tilde{Z}_{\hat{n}}^{(e)}$ in set $\mathscr{N}_l$, then set $a_{\hat{n}}=1$;
\item Reallocate the PRBs and computation resource using graph coloring method and optimization method, which are described in detail at the following subsections;
\item Recalculate the system overall overhead as in (\ref{final total system overhead});
\item If the system overall overhead $Z$ is less than that in the last iteration, set the present offloading profile $\boldsymbol{A}$ as the current offloading decision, i.e., keep $a_{\hat{n}}=1$; if not, hold the previous offloading profile as current offloading decision, i.e., restore as $a_{\hat{n}}=0$;
\item Return to the second step until all the zero elements of $\boldsymbol{A}$ have been checked. Then the current offloading decision will be the final offloading decision. And the corresponding assignments of PRBs and computational resources are the final assignments.
\end{itemize}


\subsection{Interference Graph Construction}
 The MEC server will build the interference graph using the measurements of SeNBs
 SeNBs monitor the control channel of other SeNBs, so they can receive the reference signals transmitted by their neighboring SeNBs. Then SeNBs get identifications of their neighboring SeNBs and calculate the pass loss from each of them \cite{3GPP2012ts}. Based on final offloading decision as well as SeNB measurements, the MEC server builds an interference graph where each node  stands for a SeNB and each directed edge stands for an interference status between two SeNBs. One edge between two SeNBs is established when the ratio of channel gains from the interfering SeNB to that from the serving SeNB exceeds a pre-defined threshold \cite{3GPP2012ts}.

\subsection{Normalization}
In order to allocate PRBs to UEs which will offload computation tasks, it is necessary to normalize the number of PRBs $w_n$ estimated by UE $n$ first, like in (\ref{normalization}):
\begin{equation}
\label{normalization offload}
\tilde{M}_n'=K\frac{w_n}{\sum\limits_{n \in \mathscr{N}_e}w_n},  \quad  n\in \mathscr{N}_e.
\end{equation}

But here a PRB reuse parameter $\lambda$ is introduced to achieve frequency reuse:
\begin{equation}
\label{frequency reuse}
M_n=min\left(\lfloor\lambda \tilde{M}_n'\rceil,K \right), \quad n\in \mathscr{N}_e,
\end{equation}
where
$\lfloor.\rceil$ stands for rounding to the nearest integer number. The purpose of the introduction of $\lambda$ is controlling the amount of frequency reuse. 

\subsection{Graph Coloring}
An improved graph coloring method based on \cite{Bre79} is adopted here to allocate PRBs to UEs.
In graph coloring, one color stands for one PRB, and one vertex represents a SeNB in the interference graph. With the interference graph described above, the PRB assignment problem turns into a graph coloring problem.
In order to execute graph coloring, the constructed interference graph is modified into a weighted interference graph, where the weight of every directed edge is calculated as
\begin{equation}
\label{weight of edge}
\rho_{nm}=\frac{P_n}{M_n}H_{n,m}, \quad n,m \in \mathscr{N}_e,
\end{equation}
where $H_{n,m}$ represents
the path loss from the UE associated with SeNB $n$ to SeNB $m$. 


The steps of the graph coloring PRB allocation algorithm are described below.

\subsubsection{Initialization}

In this step, the MEC server sets the PRB association table $\boldsymbol{C}$ ($N_e \times K$) mentioned above to zeros, and initializes another table, the interference table $\boldsymbol{O}$, which is also an $N_e \times K$ table.
Table $\boldsymbol{O}$ has real-valued entries $o_{nk}$ representing the sum interference from all other offloading UEs (except for its own associated UE) experienced by SeNB $n$ on PRB (color) $k$. So, $o_{nk}$ is given by
\begin{equation}
\label{interference table entry}
o_{nk}=\sum\limits_{m \in \mathscr{N}_e \backslash \{n\}} c_{mk} \frac{P_m}{M_m} H_{m,n}, \quad  n\in \mathscr{N}_e.
\end{equation}

The interference table $\boldsymbol{O}$ is set to zeros in the  initialization step. At last, a set of all the uncolored vertices $U$ is initialized as equal to the set of all offloading SeNBs (UEs) $\mathscr{N}_e$.

\subsubsection{Finding the most interfered SeNB}
\label{Finding the most interfered SeNB}
It is necessary to determine the order of SeNBs to be colored. We choose the most interfered SeNB $\bar{n}$ as the first SeNB to be colored, which is defined as the SeNB with the largest sum weights of ingoing edges:
\begin{equation}
\begin{aligned}
\label{most interfered SeNB}
\bar{n}=\operatorname*{argmax}\limits_{n \in U} \sum_{m \in \mathscr{N}_e \backslash \{n\}} \rho_{mn}
=\operatorname*{argmax}\limits_{n \in U} \sum_{m \in \mathscr{N}_e \backslash \{n\}} \frac{P_m}{M_m} H_{m,n}.
\end{aligned}
\end{equation}

If more than one SeNB have the same sum weights of ingoing edges, choose the one with the smallest $M_n$.

\subsubsection{Finding colors with the smallest interference}
In order to mitigate the interference on SeNB $\bar{n}$, the PRBs on which the smallest interference exists should be assigned to SeNB $\bar{n}$.
So it is necessary to find the colors (PRBs) with the smallest interference.
We search for these colors by looking for colors on which node (SeNB) $\bar{n}$ can achieve the highest transmission rates.
Assuming color $j$ is assigned to SeNB $\bar{n}$, we calculate the estimated rate of node $\bar{n}$ as follows:
\begin{equation}
\label{estimated rate for assigning color}
r_{\bar{n},j}=\frac{B}{K} \log_{2} \left( 1+\frac{\frac{P_{\bar{n}}}{M_{\bar{n}}}H_{\bar{n},\bar{n}}}{\sigma^2+o_{\bar{n}j}} \right),
\end{equation}
where $H_{\bar{n}}$ is the channel gain from UE $\bar{n}$ to its serving SeNB.

Now we define the estimated rate of UE $n \in \mathscr{N}_e$ under the condition that color j is assigned to UE $\bar{n}$ as follows:
\begin{equation}
\label{estimated rate for all for assigning color}
\tilde{R}_{j \rightarrow \bar{n}}^{(n)}=\sum_{q=1}^{K} \tilde{c}_{nq} r_{nq}, \quad n\in \mathscr{N}_e,
\end{equation} \\
where the notation $j\rightarrow \bar{n}$ represents that color $j$ is assigned to SeNB $\bar{n}$. The value of $\tilde{c}_{nq}$ is given by
\begin{equation}
\label{estimated PRB assignment entry}
\tilde{c}_{nq}= \left\{
\begin{aligned}
c_{nq}, \quad n \neq \bar{n} \quad or \quad q \neq j,\\
1,  \quad n=\bar{n}  \quad and \quad q=j. \\
\end{aligned}
\right.
\end{equation}
$\tilde{c}_{nq}, \forall n,q$, holds the values of the PRB association table from the previous iteration, and set the corresponding entry of presently estimated color and vertex to 1.

Next the sum of the potential rates of all the offloading SeNBs under hypothesis of assigning $j$ to $\bar{n}$ is calculated to estimate the effect of this assignment:
\begin{equation}
\label{sum potentail rate}
\tilde{S}_{j \rightarrow \bar{n}}=\sum\limits_{n \in \mathscr{N}_e} \tilde{R}_{j \rightarrow \bar{n}}^{(n)}.
\end{equation}

Then we search for the $M_{\bar{n}}$ colors which bring the largest sum rate and record them.

\subsubsection{Update Tables}

According to the PRB allocation to vertex $\bar{n}$ in the previous step, the corresponding entries of the assigned colors in table $\boldsymbol{C}$ are set to 1, and the interference caused by this new assignment is calculated and updated in table $\boldsymbol{O}$.

\subsubsection{Update the set of uncolored vertices}
\label{Update the set of uncolored vertices}
The vertex (SeNB) $\bar{n}$ got colored in this loop will be excluded from the uncolored vertices set in this step.

\subsubsection{Check whether all vertices are colored}

The uncolored vertices set $U$ will be checked. If the set $U$ is not empty, steps \ref{Finding the most interfered SeNB} to \ref{Update the set of uncolored vertices} will be repeated. If set $U$ is empty, we will go to the next step.

\subsubsection{Color assignment}

The set of colors (PRBs), which we assume is represented by $\eta_n,n\in \mathscr{N}_e$, will be allocated to the corresponding vertices (SeNBs) according to the PRB association table $\boldsymbol{C}$.

Then the achieved offloading rate of each offloading UE $n\in \mathscr{N}_e$ with the optimal PRB set assigned to it is calculated as:
\begin{equation}
\label{offloading rate with optimal PRB}
R^{(n)}=\frac{B}{K} \sum_{j \in \eta_{n}} \log_{2} \left(1+\frac{\frac{P_n}{M_{n}}H_{n,n}}{\sigma^2+\sum\limits_{m \in \mathscr{N}_e \backslash \{n\}} c_{mj} \frac{P_m}{M_{m}} H_{m,n}} \right).
\end{equation}

Based on the offloading rate, the offloading time and energy consumption of each offloading UE $n$ can be respectively given as:
\begin{equation}
\label{offloading time with optimal PRB}
T_{n,off}^{(e)}=B_n/R^{(n)}, \quad n\in \mathscr{N}_e,
\end{equation}
and
\begin{equation}
\label{offloading energy with optimal PRB}
E_n^{(e)}=\frac{P_nB_n}{R^{(n)}}, \quad n\in \mathscr{N}_e.
\end{equation}

\subsection{Computation Resource Allocation}

The computation resource of the MEC server is assigned to each offloading UE in this step. Let $F^{(e)}_n$ denote the computation resource assigned to UE $n$ ($n\in \mathscr{N}_e$). Because the energy consumption of the MEC server is not taken into consideration in our solution, here we only calculate the time consumption of computation task execution in the MEC server, for each offloading UE $n$, $T_{n,exe}^{(e)}={D_n}/{F^{(e)}_n}, \quad n\in \mathscr{N}_e$.

Then the MEC server allocates the computation resource to each offloading UE based on the following two kinds of objective functions:

\subsubsection{Minimize max-time}
The objective is to minimize the largest task execution time consumption $T_{n,exe}^{(e)}$ among all $n\in \mathscr{N}_e$. The problem can be formulated as:
\begin{equation}
\label{computation resource allocation1}
\begin{aligned}
\underset{\{F^{(e)}_n\}}{\text{min}} \quad \underset{n}{\text{max}} \quad &T_{n,exe}^{(e)} \\
\text{subject to} \quad &T_{n,exe}^{(e)}\leq T_n^{(l)}-T_{n,off}^{(e)}, \forall n\in \mathscr{N}_e \\
&\sum_{n\in \mathscr{N}_e} F^{(e)}_n=F \\
&F^{(e)}_n>0, \forall n\in \mathscr{N}_e
\end{aligned}
\end{equation}


\subsubsection{Minimize sum-time}
The objective is to minimize the total task execution time consumptions ($T_{n,exe}^{(e)}$) of all offloading UEs ($n\in \mathscr{N}_e$). The problem can be formulated as:
\begin{equation}
\label{computation resource allocation2}
\begin{aligned}
\underset{\{F^{(e)}_n\}}{\text{min}} \quad &\sum_{n\in \mathscr{N}_e} T_{n,exe}^{(e)} \\
\text{subject to} \quad &T_{n,exe}^{(e)}\leq T_n^{(l)}-T_{n,off}^{(e)}, \forall n\in \mathscr{N}_e \\
&\sum_{n\in \mathscr{N}_e} F^{(e)}_n=F \\
&F^{(e)}_n>0, \forall n\in \mathscr{N}_e
\end{aligned}
\end{equation}


The optimization problems in (\ref{computation resource allocation1}) and (\ref{computation resource allocation2}) are convex optimization problems and are easy to solve.

Now, the total time consumption of $n \in \mathscr{N}_e$ can be calculated as:
\begin{equation}
\label{final total time}
T_{n}^{(e)}=T_{n,off}^{(e)}+T_{n,exe}^{(e)}, \quad n\in \mathscr{N}_e.
\end{equation}
Then the total consumption of $n \in \mathscr{N}_e$ is given as:
\begin{equation}
\label{final total overhead}
Z_{n}^{(e)}=\gamma_n^{(T)}T_{n}^{(e)}+\gamma_n^{(E)}E_{n}^{(e)}, \quad n\in \mathscr{N}_e.
\end{equation}
Then we have the total overhead of the whole system:
\begin{equation}
\label{final total system overhead}
Z=\sum\limits_{i\in\mathscr{N}_l}Z_i^{(l)}+\sum\limits_{j\in\mathscr{N}_e} Z_{j}^{(e)},
\end{equation}
where $Z_i^{(l)}$ is given in (\ref{local total overhead}).

\subsection{Final Decision}
As described in Subsection \ref{Resource Reallocation}, when the final offloading decision vector $\boldsymbol{A}^{\ast}$ is settled, the corresponding PRB association table $\boldsymbol{C}$ and computational resource assignments $F^{(e)}_n, n\in \mathscr{N}_e$  are determined as the final assignment decisions for frequency and computation resources. Then the final overall system overhead can be calculated as in (\ref{final total system overhead}).

\section{Simulation Results and Discussions}
\label{numerical results}
In this section, simulation results of the proposed scheme are presented in comparison with several baseline schemes.
We considered 9 small cells randomly deployed in a $120 \times 120$  $m^2$ area.
The important simulation parameters employed in the simulations, unless mentioned otherwise, are summarized in table \ref{table_2}.

\begin{table}[!t]
\centering
\caption{Simulation Parameters}
\label{table_2}
\begin{tabular}{ll}
\toprule
Parameter & Value\\
\midrule
Bandwidth         & 20MHz \\
Transmission power of UE $n$, $P_n$         & 100 mWatts \\
Background noise $\sigma^2$         & -100 dBm \\
Decision weights $\gamma_n^T=\gamma_n^E$         & 0.5 \\
Data size for computation offloading $B_n$         & 420 KB \\
Total number of CPU cycles of computation task $D_n$         & 1,000 Megacycles \\
Computation capability of UE $n$, $F_n^l$         & 0.7 GHz \\
Computation capability of the MEC server $F$         & 100 GHz \\
$$         &  \\

\bottomrule
\end{tabular}
\end{table}

\begin{figure}[!t]
\centering
\includegraphics[width=0.43\textwidth]{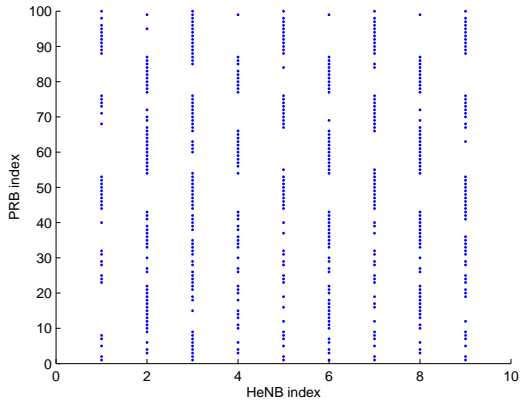}
\caption{PRB distribution among SeNBs.}
\label{PRB_allocation}
\end{figure}

\begin{figure}[!t]
\centering
\includegraphics[width=0.43\textwidth]{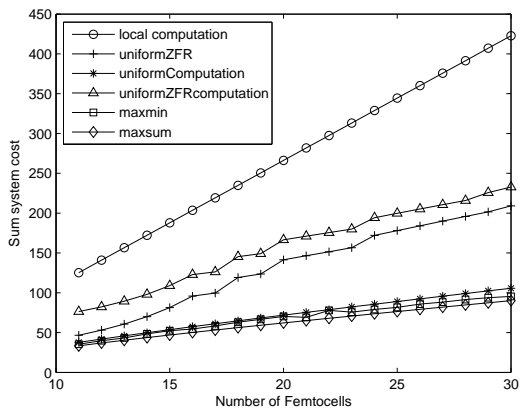}
\caption{Sum system cost versus the number of small cells.}
\label{last_figure}
\end{figure}

Fig. \ref{PRB_allocation} shows the PRB distribution among 9 SeNBs.
It can be seen that the same PRBs are reused by the SeNBs far away, rather than neighbouring SeNBs. It is because this is the most effective way that the interference among neighboring SeNBs can be mitigated. The sum of system costs of the proposed scheme and other baseline solutions in respect to the number of small cells are showed in Fig. \ref{last_figure}.
Due to the fact that the computational resource and PRBs are all allocated dynamically, and frequency reuse is allowed among small cells in the offloading process, the proposed scheme under the two objective functions achieves the lowest sum of costs among all the solutions.

\section{Conclusions}
\label{conclusion}
In this paper, we presented an integrated framework for computation offloading and interference management in heterogeneous cellular networks with MEC. We took into consideration the computation offloading decision, physical resource block allocation, and MEC computation resource allocation problems in this framework. Then, we derived the solutions to these three problems.
Simulation results demonstrate that the proposed scheme can achieve better performance than other baseline solutions under various system parameters.

\section*{Acknowledgment}
This work is jointly supported by the National Natural Science Foundation of China (Grant No. 61571073) and  the National High Technology Research and Development Program of China (Grant No. 2014AA01A701).

\bibliography{Wang_References}

\end{document}